# Kinetically Stable Single Crystals of Perovskite-Phase CsPbI$_3$

Daniel B. Straus, Shu Guo, and Robert J. Cava*

Department of Chemistry, Princeton University, Princeton, NJ 08544 USA

*Author to whom correspondence should be addressed. Email: rcava@princeton.edu

**Abstract**

We use solid-state methods to synthesize single crystals of perovskite-phase cesium lead iodide ($\gamma$-CsPbI$_3$) that are kinetically stable at room temperature. Single crystal X-ray diffraction characterization shows that the compound is orthorhombic with the GdFeO$_3$ structure at room temperature. Unlike conventional semiconductors, the optical absorption and the joint density-of-states of bulk $\gamma$-CsPbI$_3$ is greatest near the band edge and decreases beyond $E_g$ for at least 1.9 eV. Bulk $\gamma$-CsPbI$_3$ does not show an excitonic resonance and has an optical band gap of 1.63(3) eV, ~90 meV smaller than has been reported in thin films; these and other differences indicate that previously-measured thin film $\gamma$-CsPbI$_3$ shows signatures of quantum confinement. By flowing gases over $\gamma$-CsPbI$_3$ during *in situ* powder X-ray diffraction measurements, we confirm that $\gamma$-CsPbI$_3$ is stable to oxygen but rapidly and catalytically converts to non-perovskite $\delta$-CsPbI$_3$ in the presence of moisture. Our results on bulk $\gamma$-CsPbI$_3$ provide vital parameters for theoretical and experimental investigations into perovskite-phase CsPbI$_3$ that will the guide the design and synthesis of atmospherically stable inorganic halide perovskites.

Halide perovskites have attracted tremendous interest due to their promise as a solution-processable active layer in solar cells with efficiencies rivaling commercial silicon solar cells.[1,2] Halide perovskites have the formula $AMX_3$, where A is a small organic (e.g. $CH_3NH_3^+$) or $Cs^+$ cation, M is typically $Pb^{2+}$ or $Sn^{2+}$, and X is a halide. The highest efficiency halide perovskite solar cells incorporate methylammonium lead iodide ($CH_3NH_3PbI_3$), which tends to decompose under normal operating conditions.[3]

One approach to stabilizing halide perovskite solar cells is to replace the organic cation with an inorganic $Cs^+$ cation.[4] While cesium lead iodide ($CsPbI_3$) has a much higher decomposition temperature than $CH_3NH_3PbI_3$,[5] introducing $Cs^+$ is problematic because the size of the $Cs^+$ cation is near the lower limit for forming a lead iodide perovskite.[6,7] Accordingly, $CsPbI_3$ is only thermodynamically stable as a perovskite above ~325 °C,[8] where it is cubic ($\alpha$-$CsPbI_3$).[5] At ambient conditions it rapidly converts to a yellow material consisting of one-dimensional chains of Pb-I octahedra ($\delta$-$CsPbI_3$).[9,10] Moisture is reported to accelerate conversion from perovskite-phase $CsPbI_3$ to $\delta$-$CsPbI_3$,[8,11] and perovskite-phase $CsPbI_3$ is unstable in atmosphere and unsuitable for solar cells without additional stabilization. Small grain sizes stabilize perovskite-phase $CsPbI_3$,[4,12,13] and atmospherically-stable perovskite-phase $CsPbI_3$ has been synthesized as quantum dots[13] and thin films with grain sizes of ~100nm,[12] allowing for the creation of $CsPbI_3$-based solar cells that function in atmosphere.[13–15]

To the best of our knowledge, all studies to-date of perovskite-phase $CsPbI_3$ have been on thin-films, powders, or nanocrystals.[16–18] It is widely held that the large 6.9% volume change from $\delta$-$CsPbI_3$ to cubic perovskite-phase $\alpha$-$CsPbI_3$ would cause single crystals to fracture, rendering the synthesis of single crystals of perovskite-phase $CsPbI_3$ impossible.[5]

Here we directly synthesize and characterize macroscopic single crystals of perovskite-phase CsPbI$_3$ that are kinetically stable at room temperature in the absence of moisture. Using single-crystal X-ray diffraction (SCXRD) measurements, we confirm recent powder X-ray diffraction (PXRD) refinements[16,17] that show perovskite-phase CsPbI$_3$ is orthorhombic at room temperature ($\gamma$-CsPbI$_3$) adopting the GdFeO$_3$ structure. We find a smaller band gap $E_g$ of 1.63(3) eV for perovskite-phase $\gamma$-CsPbI$_3$ than previously reported and show that its strongest absorption between $E_g$ and 3.5 eV is near the band edge, indicating that previous reports of the optical character of $\gamma$-CsPbI$_3$ are for quantum-confined material. By monitoring the optical absorption during the conversion from $\gamma$-CsPbI$_3$ to $\delta$-CsPbI$_3$, we demonstrate that quantum-confined $\gamma$-CsPbI$_3$ shows enhanced stability, supporting theoretical and experimental findings that small grain sizes stabilize $\gamma$-CsPbI$_3$.[4,12,13] Lastly, by flowing gases over the sample *in situ* during PXRD experiments, we demonstrate that while perovskite-phase $\gamma$-CsPbI$_3$ is stable in dry argon and dry oxygen atmospheres, it rapidly and completely converts to $\delta$-CsPbI$_3$ upon exposure to humid argon. Our structural and optical data provide fundamental parameters necessary for theoretical studies on CsPbI$_3$ that will inform the design and synthesis of stable all-inorganic halide perovskites.

We synthesize CsPbI$_3$ single crystals *via* a solid-state method. CsPbI$_3$ melts congruently at ~480 °C,[19] and we melt a stoichiometric mixture of dry CsI and PbI$_2$ sealed in an evacuated ampoule. The black perovskite phase is synthesized by slowly cooling the melt from 550 °C to 370 °C followed by rapid quenching in an ice water bath. Figure 1A shows bulk perovskite-phase $\gamma$-CsPbI$_3$ in an evacuated ampoule 39 days after synthesis. Perovskite-phase CsPbI$_3$ is known to be unstable in air,[4,8,11] so extreme care is taken to minimize air exposure.

At room temperature, perovskite-phase γ-CsPbI$_3$ is orthorhombic[16,17] in the *Pnma* space group with the GdFeO$_3$ structure type, which following convention we refer to as γ-CsPbI$_3$. Structural distortion in perovskites is categorized by Glazer,[20] and γ-CsPbI$_3$ follows tilt system #10 with three octahedral tilt angles, two of which are identical. We find tilt angles of 14.00(2)° and 10.08(3)°. The PbI$_4^{2-}$ octahedra are distorted with I-Pb-I bond angles of 89.29(3)°, 88.44(3)° and 89.117(9)° and show more distortion than CH$_3$NH$_3$PbI$_3$ which is tetragonal at room temperature with I-Pb-I bond angles of 88.94(13), 89.980(5) and tilt angles of 1.1(2) and 8.23(3).[9] The distortion in γ-CsPbI$_3$ may cause its thermodynamic instability.

Our 295 K SCXRD structure for perovskite-phase γ-CsPbI$_3$ is depicted in Figure 1B and described in Tables 1, S1, and S2. While our structure is similar to two recent structures found by Rietveld refinements of PXRD patterns,[16,17] characterization *via* SCXRD avoids complications due to overlapping reflections, preferred orientation, and background subtraction of the air-free holder that are inherent to PXRD pattern refinements.[16,17] Our single crystal refinement provides reliable structural parameters necessary for accurate theoretical calculations.

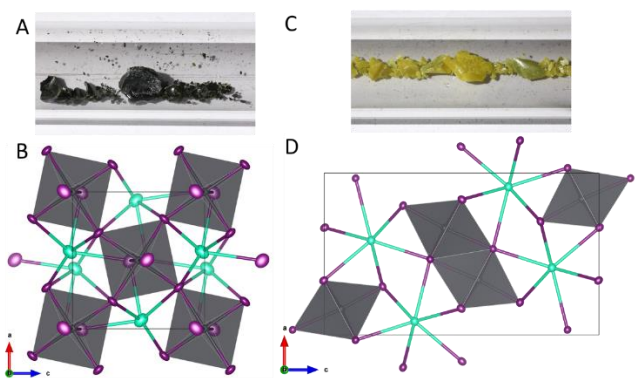

Figure 1: (A) Photograph of γ-CsPbI$_3$. (B) SCXRD structure of γ-CsPbI$_3$. (C) Photograph of δ-CsPbI$_3$. (D) SCXRD structure of δ-CsPbI$_3$. (B) and (D) visualized using VESTA[21] with Cs in green, Pb in grey, and I in purple, all depicted as 50% thermal ellipsoids.

Table 1: Crystallographic structural parameters

|  | $\gamma$-CsPbI$_3$ | $\delta$-CsPbI$_3$ |
|---|---|---|
| Empirical formula | CsI$_3$Pb | CsI$_3$Pb |
| Formula weight | 720.80 | 720.80 |
| Temperature/K | 295 | 295 |
| Crystal system | orthorhombic | orthorhombic |
| Space group | *Pnma* | *Pnma* |
| a/Å | 8.8637(8) | 10.4500(5) |
| b/Å | 12.4838(12) | 4.7965(2) |
| c/Å | 8.5783(8) | 17.7602(8) |
| Volume/Å$^3$ | 949.21(15) | 890.20(7) |
| Z | 4 | 4 |
| $\rho_{calc}$g/cm$^3$ | 5.044 | 5.378 |
| μ/mm$^{-1}$ | 31.213 | 33.282 |
| Crystal size/mm$^3$ | 0.053 × 0.047 × 0.029 | 0.038 × 0.036 × 0.017 |
| Crystal color | black | yellow |
| Goodness-of-fit on F$^2$ | 1.047 | 1.107 |
| Final R indexes [I>=2σ (I)] | R$_1$ = 0.0349, wR$_2$ = 0.0599 | R$_1$ = 0.0202, wR$_2$ = 0.0313 |
| Final R indexes [all data] | R$_1$ = 0.0645, wR$_2$ = 0.0679 | R$_1$ = 0.0290, wR$_2$ = 0.0330 |

If CsPbI$_3$ is slowly cooled to room temperature from 550 °C, yellow $\delta$-CsPbI$_3$ forms (Figure 1C), confirming $\delta$-CsPbI$_3$ is the thermodynamic product and $\gamma$-CsPbI$_3$ is the kinetic product. The 6.9% volume change from to the cubic perovskite phase $\alpha$-CsPbI$_3$ to $\delta$-CsPbI$_3$ has been hypothesized to cause single crystals to fracture,[5] but we find single crystals of yellow $\delta$-CsPbI$_3$ in the slow-cooled material. Our 295 K SCXRD structure is shown in Figure 1D and described in Tables 1, S1, and S2. Like $\gamma$-CsPbI$_3$, $\delta$-CsPbI$_3$ is also orthorhombic and in the *Pnma* space group but with a non-perovskite NH$_4$CdCl$_3$ structure type.[10] Despite the platy crystal habit of our crystals, our crystal structure does not significantly differ from a reported SCXRD structure on needle habit $\delta$-CsPbI$_3$ crystals synthesized in aqueous hydriodic acid.[9]

Figure 2A shows the absorption spectra of bulk $\gamma$-CsPbI$_3$ (black) and $\delta$-CsPbI$_3$ (yellow) converted from diffuse reflectance spectra (Figure S1). Raman scattering spectra are shown in Figure S2. Using the direct band gap Tauc formalism for allowed transitions,[22] we find the band gap of δ-CsPbI$_3$ synthesized through solid-state methods is 2.58(4) eV (Figure S3). By comparing the heights of the absorption peaks, we estimate that the band edge absorption cross-section of γ-CsPbI$_3$ is 3.2x larger than that of δ-CsPbI$_3$. The Tauc band gap of γ-CsPbI$_3$ is 1.63(3) eV (Figure S3), 90 meV smaller than the ~1.72 eV band gap reported for γ-CsPbI$_3$ thin films synthesized by heating δ-CsPbI$_3$.[4,8,16] Strangely, we find that after reaching a maximum near the band edge, the absorbance of γ-CsPbI$_3$ decreases with increasing energy to 3.5 eV indicating that the greatest joint density-of-states is near the band edge, in contrast with traditional direct band gap semiconductors.[23] Our absorption spectrum for γ-CsPbI$_3$ qualitatively matches a recent calculated absorption spectrum for cubic α-CsPbI$_3$[24] but differs from previous measurements on thin films of γ-CsPbI$_3$, which show an absorption coefficient increasing with energy and an excitonic absorption resonance.[8,16] In addition, smaller grain thin films show a greater increase in absorption at higher energies.[16] The shape of the absorption spectra and larger band gap for the thin film material are signatures that thin film γ-CsPbI$_3$ exhibits quantum confinement effects.[25]

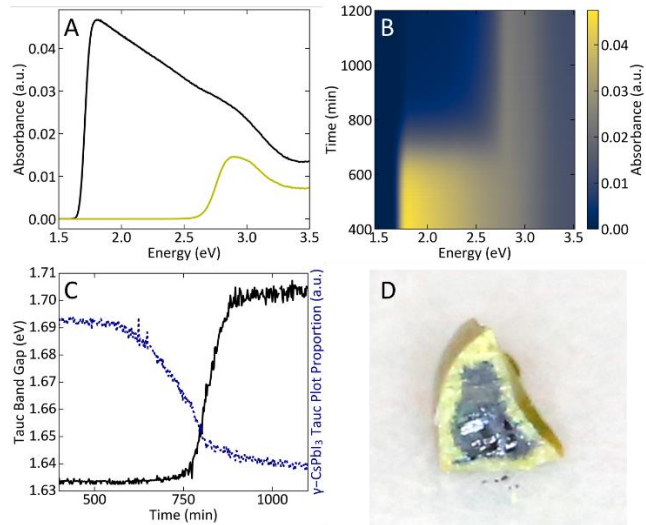

Figure 2: (A) Optical absorption spectra of (black) γ-CsPbI$_3$ and (yellow) δ-CsPbI$_3$. (B) Absorption spectra of γ-CsPbI$_3$ converting to δ-CsPbI$_3$, with (C) the band gap and fraction of Tauc absorbance of γ-CsPbI$_3$. (D) Cross-section of a piece of γ-CsPbI$_3$ after 6 minutes in air.

We optically monitor the conversion of perovskite-phase $\gamma$-CsPbI$_3$ to $\delta$-CsPbI$_3$ (Figure 2B) by incompletely sealing $\gamma$-CsPbI$_3$ in a diffuse reflectance powder cell. Over time, the absorbance of $\gamma$-CsPbI$_3$ decreases and its band gap increases to 1.70 eV (Figure 2C) while $\delta$-CsPbI$_3$ simultaneously appears in the absorption spectrum due to the reaction of $\gamma$-CsPbI$_3$ with the atmosphere. In addition, like previously reported spectra of thin film $\gamma$-CsPbI$_3$ but unlike bulk $\gamma$-CsPbI$_3$, the absorption spectrum of the partially converted, wider band gap $\gamma$-CsPbI$_3$ (orange, Figure S4) does not show decreasing absorption at higher energies. By leaving a piece of bulk $\gamma$-CsPbI$_3$ in air for 6 minutes and then cross-sectioning it, we observe that the conversion to $\delta$-CsPbI$_3$ travels inwards from the outer surfaces (Figure 2D). These findings further support our inference that the previously reported 1.72 eV band gap of $\gamma$-CsPbI$_3$ is caused by quantum confinement effects. Finally, the slower conversion of $\gamma$-CsPbI$_3$ to $\delta$-CsPbI$_3$ as the size of $\gamma$-CsPbI$_3$ particles decreases (Figure 2B-C) supports that small grain sizes help to stabilize $\gamma$-CsPbI$_3$.[4,12]

Humidity has been reported to convert perovskite-phase $\gamma$-CsPbI$_3$ to $\delta$-CsPbI$_3$,[8,11] and our synthesis of kinetically stable bulk $\gamma$-CsPbI$_3$ requires the use of rigorously dry reagents. The sample shown in Figure 3A is made under identical conditions as the sample in Figure 1A but uses reagents that are not completely anhydrous—black $\gamma$-CsPbI$_3$ slowly turns to yellow $\delta$-CsPbI$_3$ despite being synthesized in an evacuated, nominally atmosphere-free ampoule. The stability of the strictly anhydrous material in the evacuated tube indicates that the relatively small amount of water vapor from the not-strictly-anhydrous starting materials has caused the $\gamma$-CsPbI$_3$ to $\delta$-CsPbI$_3$ conversion. The small amount of water present compared to the amount of material and the continuing $\gamma$ to $\delta$ conversion indicates that the role of water vapor is catalytic in accord with a previous report,[8] and that the yellow $\delta$-CsPbI$_3$ material is not a hydrate, in accord with the similarity of our $\delta$-CsPbI$_3$

SCXRD structure to a previous structure on crystals synthesized in aqueous hydriodic acid.[9] In contrast, we recently showed that what was originally reported as $Cs_2PdCl_4$ is actually a hydrate.[26]

To further investigate the effect of water vapor on $\gamma$-$CsPbI_3$, we perform PXRD while flowing gas *in situ* over ground powders of $\gamma$-$CsPbI_3$. Figure 3B shows the PXRD pattern of $\gamma$-$CsPbI_3$ under flowing dry argon with a Rietveld refinement to our SCXRD structure shown in red. Flowing dry oxygen over the sample does not result in any changes (Figure 3C). However, flowing argon bubbled through deionized water over the sample to create a water-saturated, oxygen and carbon dioxide-free ambient results in complete conversion to $\delta$-$CsPbI_3$ within 1 minute (Figure 3C). Figure 3D shows the PXRD pattern of the converted $\gamma$-$CsPbI_3$ with a refinement to our SCXRD structure shown in red, confirming complete conversion to $\delta$-$CsPbI_3$ and demonstrating the extreme moisture sensitivity of bulk $\gamma$-$CsPbI_3$.

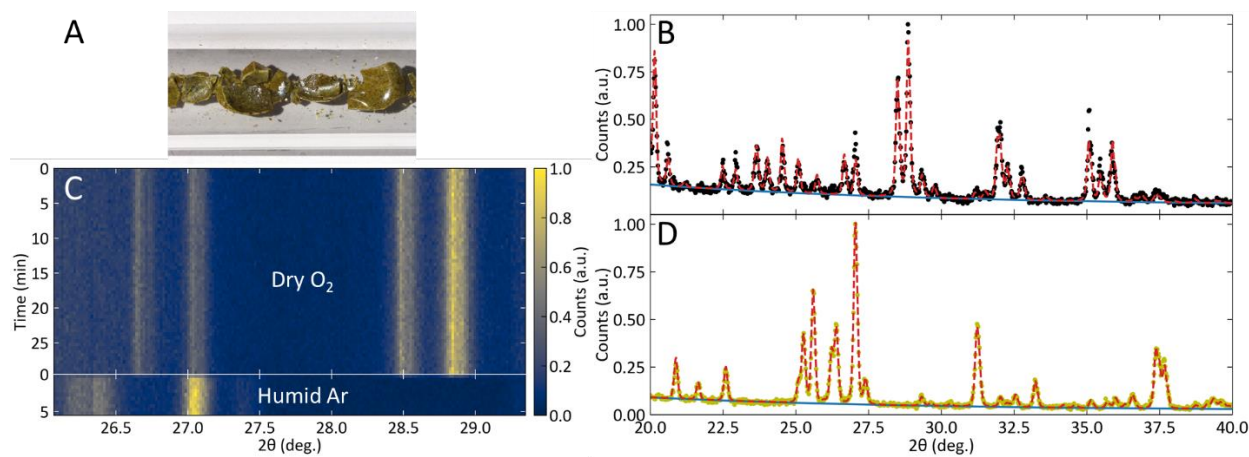

Figure 3: (A) Photograph of $\gamma$-$CsPbI_3$ in evacuated ampoule made using non-anhydrous reagents 39 days after synthesis. (B) (black) PXRD pattern of $\gamma$-$CsPbI_3$ under flowing dry argon, with (red) Rietveld refinement and (blue) background. (C) PXRD patterns of $\gamma$-$CsPbI_3$ under flowing dry oxygen and humid argon. (D) (yellow) PXRD pattern of $\delta$-$CsPbI_3$ formed by flowing humid argon over $\gamma$-$CsPbI_3$ with (red) Rietveld refinement and (blue) background.

We demonstrate that bulk single crystals of perovskite-phase γ-CsPbI$_3$ are kinetically stable at room temperature in inert conditions despite their thermodynamic instability. We find that the band gap of bulk γ-CsPbI$_3$ is 1.63(3) eV, ~90 meV lower than the band gap of thin film γ-CsPbI$_3$. In addition, unlike thin film γ-CsPbI$_3$, bulk γ-CsPbI$_3$ does not have a distinct excitonic resonance, and these differences suggest that thin film γ-CsPbI$_3$ measured in previous studies[8,16] exhibits quantum confinement effects. Optical absorption measurements show that the greatest joint density-of-states is near the band edge and decreases with increasing energy to at least 1.9 eV beyond the band edge, in contrast to conventional semiconductors. Through synthetic variation and in-situ PXRD measurements, we illustrate that water vapor catalyzes the conversion of γ-CsPbI$_3$ to δ-CsPbI$_3$, and our SCXRD structure and optical characterization of bulk γ-CsPbI$_3$ provide detailed parameters that can be used for more accurate theoretical modelling of the properties, stability, and degradation mechanisms of γ-CsPbI$_3$. Our results are vital for theoretical and experimental studies on perovskite-phase CsPbI$_3$ given that its bulk properties are differ from what is observed for polycrystalline thin films and will help guide the design and synthesis of atmospherically stable inorganic halide perovskites.

**Acknowledgments**

The synthesis and general characterization analysis of the compound was supported by the Gordon and Betty Moore Foundation, grant GBMF-4412. The crystal structure refinement was supported by the US Department of Energy, Division of Basic Energy Sciences, grant DE-SC0019331.

**References**

(1) De Angelis, F. Celebrating 10 Years of Perovskite Photovoltaics. *ACS Energy Lett.* **2019**, *4*, 853–854.

(2) NREL. Best Research-Cell Efficiencies http://www.nrel.gov/pv/assets/images/efficiency-chart.png (accessed May 23, 2019).

(3) Conings, B.; Drijkoningen, J.; Gauquelin, N.; Babayigit, A.; D'Haen, J.; D'Olieslaeger, L.; Ethirajan, A.; Verbeeck, J.; Manca, J.; Mosconi, E.; et al. Intrinsic Thermal Instability of Methylammonium Lead Trihalide Perovskite. *Adv. Energy Mater.* **2015**, *5*, 1500477.

(4) Eperon, G. E.; Paternò, G. M.; Sutton, R. J.; Zampetti, A.; Haghighirad, A. A.; Cacialli, F.; Snaith, H. J. Inorganic Caesium Lead Iodide Perovskite Solar Cells. *J. Mater. Chem. A* **2015**, *3*, 19688–19695.

(5) Trots, D. M.; Myagkota, S. V. High-Temperature Structural Evolution of Caesium and Rubidium Triiodoplumbates. *J. Phys. Chem. Solids* **2008**, *69*, 2520–2526.

(6) Goldschmidt, V. M. Crystal Structure and Chemical Constitution. *Trans. Faraday Soc.* **1929**, *25*, 253.

(7) Travis, W.; Glover, E. N. K.; Bronstein, H.; Scanlon, D. O.; Palgrave, R. G. On the Application of the Tolerance Factor to Inorganic and Hybrid Halide Perovskites: A Revised System. *Chem. Sci.* **2016**, *7*, 4548–4556.

(8) Dastidar, S.; Hawley, C. J.; Dillon, A. D.; Gutierrez-Perez, A. D.; Spanier, J. E.; Fafarman, A. T. Quantitative Phase-Change Thermodynamics and Metastability of Perovskite-Phase Cesium Lead Iodide. *J. Phys. Chem. Lett.* **2017**, *8*, 1278–1282.

(9)   Stoumpos, C. C.; Malliakas, C. D.; Kanatzidis, M. G. Semiconducting Tin and Lead Iodide Perovskites with Organic Cations: Phase Transitions, High Mobilities, and Near-Infrared Photoluminescent Properties. *Inorg. Chem.* **2013**, *52*, 9019–9038.

(10)  Møller, C. Crystal Structure and Photoconductivity of Cæsium Plumbohalides. *Nature* **1958**, *182*, 1436–1436.

(11)  Moller, C. K. The Structure of Caesium Plumbo Iodide CsPbI3. *Mater. Fys. Medd. Dan. Vid. Selsk.* **1959**, *32*, 1–18.

(12)  Zhao, B.; Jin, S. F.; Huang, S.; Liu, N.; Ma, J. Y.; Xue, D. J.; Han, Q.; Ding, J.; Ge, Q. Q.; Feng, Y.; et al. Thermodynamically Stable Orthorhombic γ-CsPbI3 Thin Films for High-Performance Photovoltaics. *J. Am. Chem. Soc.* **2018**, *140*, 11716–11725.

(13)  Swarnkar, A.; Marshall, A. R.; Sanehira, E. M.; Chernomordik, B. D.; Moore, D. T.; Christians, J. A.; Chakrabarti, T.; Luther, J. M. Quantum Dot-Induced Phase Stabilization of -CsPbI3 Perovskite for High-Efficiency Photovoltaics. *Science* **2016**, *354*, 92–95.

(14)  Sanehira, E. M.; Marshall, A. R.; Christians, J. A.; Harvey, S. P.; Ciesielski, P. N.; Wheeler, L. M.; Schulz, P.; Lin, L. Y.; Beard, M. C.; Luther, J. M. Enhanced Mobility CsPbI 3 Quantum Dot Arrays for Record-Efficiency, High-Voltage Photovoltaic Cells. *Sci. Adv.* **2017**, *3*, eaao4204.

(15)  Luo, P.; Xia, W.; Zhou, S.; Sun, L.; Cheng, J.; Xu, C.; Lu, Y. Solvent Engineering for Ambient-Air-Processed, Phase-Stable CsPbI3 in Perovskite Solar Cells. *J. Phys. Chem. Lett.* **2016**, *7*, 3603–3608.

(16)  Sutton, R. J.; Filip, M. R.; Haghighirad, A. A.; Sakai, N.; Wenger, B.; Giustino, F.; Snaith,

H. J. Cubic or Orthorhombic? Revealing the Crystal Structure of Metastable Black-Phase CsPbI 3 by Theory and Experiment. *ACS Energy Lett.* **2018**, *3*, 1787–1794.

(17) Marronnier, A.; Roma, G.; Boyer-Richard, S.; Pedesseau, L.; Jancu, J.-M.; Bonnassieux, Y.; Katan, C.; Stoumpos, C. C.; Kanatzidis, M. G.; Even, J. Anharmonicity and Disorder in the Black Phases of Cesium Lead Iodide Used for Stable Inorganic Perovskite Solar Cells. *ACS Nano* **2018**, *12*, 3477–3486.

(18) Stoumpos, C. C.; Kanatzidis, M. G. The Renaissance of Halide Perovskites and Their Evolution as Emerging Semiconductors. *Acc. Chem. Res.* **2015**, *48*, 2791–2802.

(19) Sharma, S.; Weiden, N.; Weiss, A. Phase Diagrams of Quasibinary Systems of the Type: ABX 3 — A′BX 3 ; ABX 3 — AB′X 3 , and ABX 3 — ABX′ 3 ; X = Halogen. *Zeitschrift für Phys. Chemie* **1992**, *175*, 63–80.

(20) Glazer, A. M. Simple Ways of Determining Perovskite Structures. *Acta Crystallogr. Sect. A* **1975**, *31*, 756–762.

(21) Momma, K.; Izumi, F. VESTA 3 for Three-Dimensional Visualization of Crystal, Volumetric and Morphology Data. *J. Appl. Crystallogr.* **2011**, *44*, 1272–1276.

(22) Viezbicke, B. D.; Patel, S.; Davis, B. E.; Birnie, D. P. Evaluation of the Tauc Method for Optical Absorption Edge Determination: ZnO Thin Films as a Model System. *Phys. status solidi* **2015**, *252*, 1700–1710.

(23) Sze, S. M.; Ng, K. K. *Physics of Semiconductor Devices*, 3rd ed.; Hoboken, NJ, 2007.

(24) Afsari, M.; Boochani, A.; Hantezadeh, M. Electronic, Optical and Elastic Properties of Cubic Perovskite CsPbI3: Using First Principles Study. *Optik (Stuttg).* **2016**, *127*, 11433–

11443.

(25) Wang, Y.; Herron, N. Nanometer-Sized Semiconductor Clusters: Materials Synthesis, Quantum Size Effects, and Photophysical Properties. *J. Phys. Chem.* **1991**, *95*, 525–532.

(26) Ni, D.; Guo, S.; Yang, Z. S.; Kuo, H.-Y.; Cevallos, F. A.; Cava, R. J. A Monoclinic Form of Anhydrous $Cs_2PdCl_4$. *Solid State Sci.* **2019**, *87*, 118–123.

# Supporting Information: Kinetically Stable Single Crystals of Perovskite-Phase CsPbI$_3$

Daniel B. Straus, Shu Guo, and Robert J. Cava*

Department of Chemistry, Princeton University, Princeton, NJ 08544 USA

*Author to whom correspondence should be addressed. Email: rcava@princeton.edu

**Methods**

A stoichiometric ratio of PbI$_2$ and CsI are flame-sealed in a triple argon-flushed evacuated (~10 mTorr) quartz tube and heated at 550 °C for several hours, forming a dark purple melt. To form γ-CsPbI$_3$, the melt is slowly cooled at a rate of 2-9 °C to 370 °C and then quickly quenched in an ice-water bath, resulting in a shiny black solid. To directly synthesize δ-CsPbI$_3$, the melt is cooled to room temperature at a rate of 2 to 4 °C. Kinetically stable γ-CsPbI$_3$ only forms when dry PbI$_2$ and CsI are used; we use ultra-dry PbI$_2$ (Alfa-Aesar, 99.999%) either without further purification or further purified by vapor transport, and either anhydrous CsI (Sigma-Aldrich, 99.999%) without further purification or CsI (Alfa Aesar, 99.999%) that is dried by placing the CsI in a quartz tube, flame melted under dynamic vacuum, sealed in an evacuated quartz ampoule with a piece of graphite to absorb residual moisture, heated to 650 °C overnight, and subsequently cooled to room temperature. The not-completely anhydrous reagents used to synthesize the γ-CsPbI$_3$ shown in Figure 3A are CsI (Alfa Aesar, 99.999%) and PbI$_2$ (Alfa Aesar, 99.9985%) used without additional purification. All reagents are stored in an argon glove box with O$_2$ and H$_2$O levels ≤ 0.1 ppm

To acquire SCXRD data on γ-CsPbI$_3$, an evacuated quartz ampoule containing γ-CsPbI$_3$ synthesized using dry reagents is opened in an argon glove box. γ-CsPbI$_3$ is placed in degassed Parabar 10312 oil and removed from the glove box in a sealed vial. When selecting crystals for SCXRD, dry nitrogen flows over the crystals and an Oxford Cryostream flows dry nitrogen at 295 K over the sample while it is measured on the diffractometer.

SCXRD data on are collected at 295 K on γ-CsPbI$_3$ with a Bruker Kappa Apex2 CCD diffractometer and on δ-CsPbI$_3$ with a Bruker D8 Venture diffractometer equipped with a Photon 100 CMOS detector. Dry nitrogen at 295 K flows over both samples during collection. Graphite-monochromated Mo-Ka radiation (λ = 0.71073 Å) is used. The raw data are corrected for background, polarization, and the Lorentz factor and multi-scan absorption corrections are applied. The structures are analyzed by the Intrinsic Phasing method provided by the ShelXT structure solution program[1] and refined using the ShelXL least-squares refinement package with the Olex2 program.[2,3] The ADDSYM algorithm in PLATON is used to double check for possible higher symmetry.[4]

UV-Visible diffuse-reflectance spectra are collected in an Agilent Cary 5000 spectrometer using an Agilent Internal DRA-2500 diffuse reflectance accessory on powders diluted with dry MgO to 10% w/w. Dry MgO is used as the reflectance standard. Scans are taken every 1 nm with a 0.2 second integration time and a spectral bandwidth of 2 nm. For the data shown in Figure 2B, scans are taken every 2 minutes. Diffuse reflectance spectra are converted to absorption spectra using the Kubelka-Munk function. Band gaps are calculated using the algorithm in Ref.[5]. Raman spectra are taken on a Thermo-Fisher DXR Smart Raman instrument with a 780 nm HP laser. Powder X-ray diffraction (PXRD) patterns are taken on a Bruker D8 Advance Eco with a Lynxeye 1D detector. A Bruker air-free sample holder is modified to flow gases over the powder in-situ (Figure S5). Rietveld refinements are performed using Topas 5 using the SCXRD structures reported herein with lattice parameters fixed and instrumental corrections allowed to vary.

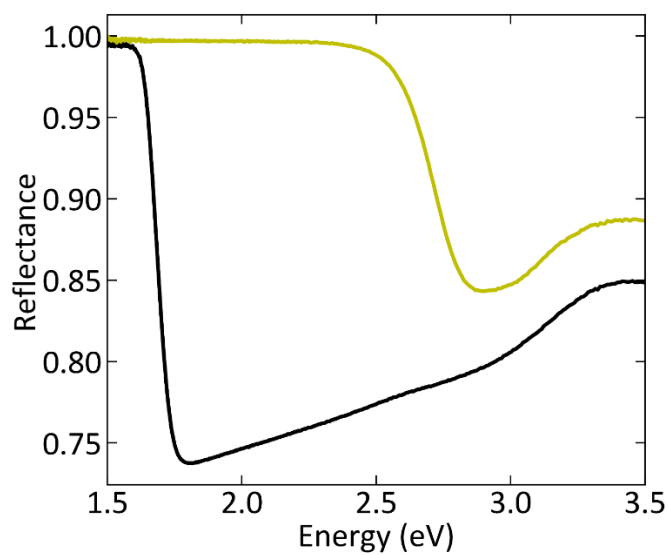

Figure S1: Reflectance spectra of (black) γ-CsPbI$_3$ and (yellow) δ-CsPbI$_3$.

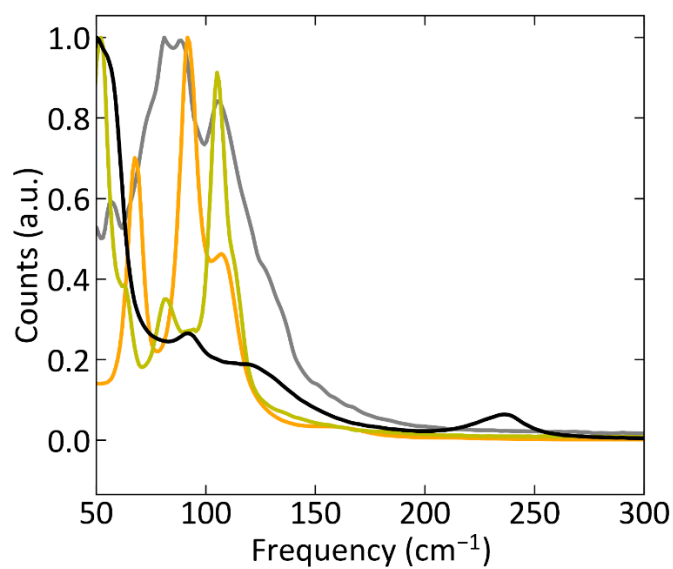

Figure S2: Raman scattering spectra for (black) γ-CsPbI$_3$, (yellow) δ-CsPbI$_3$, (grey) CsI, and (orange) PbI$_2$. γ-CsPbI$_3$ shows Raman scattering to about 150 cm$^{-1}$, with an additional peak at ~240 cm$^{-1}$ that is likely a second order replica caused by quasi-resonant 1.59 eV excitation.[6,7]

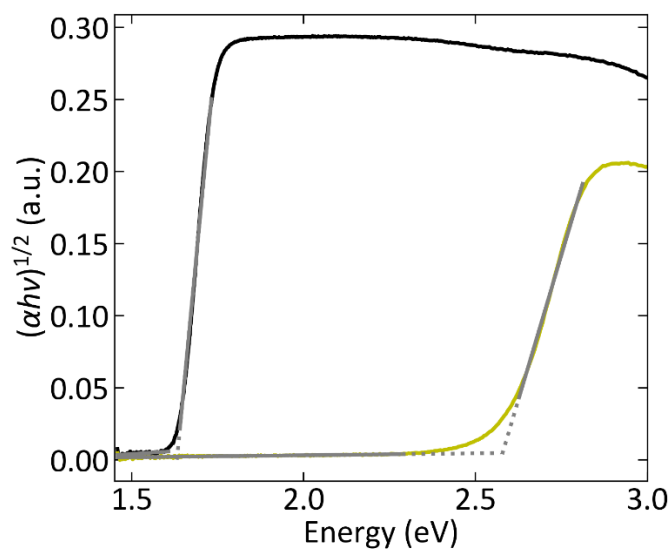

Figure S3: Direct band gap allowed transition Tauc plots of (black) γ-CsPbI$_3$ and (yellow) δ-CsPbI$_3$.

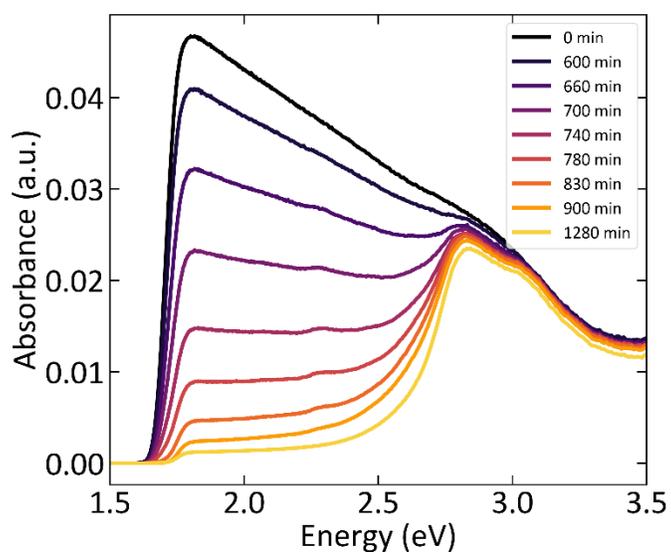

Figure S4: Select absorption spectra from Figure 2B.

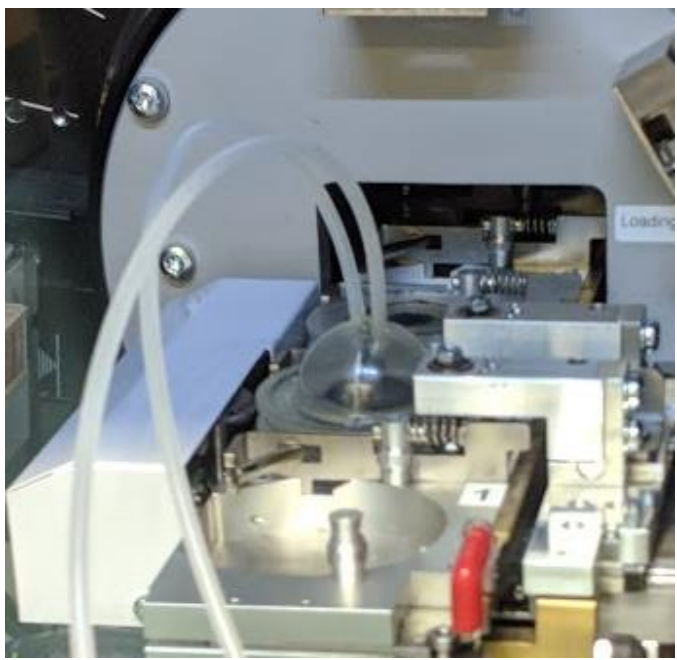

Figure S5: Sample holder for in-situ gas flow during PXRD experiments.

Table S1: collection parameters for SCXRD structures

| | | |
|---|---|---|
| F(000) | 1184 | 1184 |
| Radiation | MoKα (λ = 0.71073) | MoKα (λ = 0.71073) |
| 2Θ range for data collection/° | 5.76 to 56.56 | 6.02 to 55.042 |
| Index ranges | -11 ≤ h ≤ 11, -16 ≤ k ≤ 16, -9 ≤ l ≤ 11 | -13 ≤ h ≤ 13, -6 ≤ k ≤ 6, -23 ≤ l ≤ 23 |
| Reflections collected | 7193 | 31390 |
| Independent reflections | 1231 [$R_{int}$ = 0.0541, $R_{sigma}$ = 0.0478] | 1145 [$R_{int}$ = 0.0591, $R_{sigma}$ = 0.0173] |
| Data/restraints/parameters | 1231/0/28 | 1145/0/31 |
| Largest diff. peak/hole / e Å$^{-3}$ | 1.61/-1.84 | 0.95/-0.65 |

Table S2: Fractional atomic coordinates (Å) and equivalent isotropic displacement parameters (Å$^2$)

| | γ-CsPbI$_3$ | | | | | δ-CsPbI$_3$ | | | |
|---|---|---|---|---|---|---|---|---|---|
| Atom | x | y | z | U(eq) | Atom | x | y | z | U(eq) |
| Pb1 | 0 | 0 | 0 | 0.03112(13) | Pb1 | 0.66034(2) | 0.75000 | 0.43800(2) | 0.02803(8) |
| Cs1 | 0.43839(17) | 0.25000 | 0.02006(17) | 0.0786(4) | Cs1 | 0.58430(4) | 0.25000 | 0.17092(2) | 0.03520(11) |
| I1 | 0.50303(16) | 0.25000 | 0.56458(16) | 0.0700(4) | I1 | 0.46809(4) | 0.25000 | 0.38549(2) | 0.02938(11) |
| I2 | 0.19871(9) | 0.03272(7) | 0.30445(9) | 0.0550(2) | I2 | 0.79913(4) | 0.75000 | 0.28721(2) | 0.03251(11) |
| | | | | | I3 | 0.83673(4) | 0.25000 | 0.50160(2) | 0.03143(11) |

**References**

(1) Sheldrick, G. M. SHELXT – Integrated Space-Group and Crystal-Structure Determination. *Acta Crystallogr. Sect. A Found. Adv.* **2015**, *71*, 3–8.

(2) Sheldrick, G. M. Crystal Structure Refinement with SHELXL . *Acta Crystallogr. Sect. C Struct. Chem.* **2015**, *71*, 3–8.

(3) Dolomanov, O. V.; Bourhis, L. J.; Gildea, R. J.; Howard, J. A. K.; Puschmann, H. OLEX2 : A Complete Structure Solution, Refinement and Analysis Program. *J. Appl. Crystallogr.* **2009**, *42*, 339–341.

(4) Spek, A. L. Single-Crystal Structure Validation with the Program PLATON . *J. Appl. Crystallogr.* **2003**, *36*, 7–13.

(5) Suram, S. K.; Newhouse, P. F.; Gregoire, J. M. High Throughput Light Absorber Discovery, Part 1: An Algorithm for Automated Tauc Analysis. *ACS Comb. Sci.* **2016**, *18*, 673–681.

(6) Nakashima, S. Raman Study of Polytypism in Vapor-Grown PbI2. *Solid State Commun.* **1975**, *16*, 1059–1062.

(7) Scott, J. F.; Leite, R. C. C.; Damen, T. C. Resonant Raman Effect in Semiconductors. *Phys. Rev.* **1969**, *188*, 1285–1290.